%% file: paper.tex
\documentstyle[12pt,epsf, epic, eepic]{article}
\begin{document}

\begin{flushright}
IMSc/2000/07/37
\end{flushright}

\vskip 1cm

\begin{center}
{\Large \bf Quasi-normal modes of AdS black holes : A superpotential approach}

\vskip 1cm

{\large  T.R.Govindarajan  \footnote{e-mail: trg@imsc.ernet.in}
and V.Suneeta \footnote{ e-mail: suneeta@imsc.ernet.in}}

\vskip 1cm

{\small \it The Institute of Mathematical Sciences,
CIT Campus, Chennai 600113, India}\\

\date{\today}
\end{center}

\vskip 2cm
   
\begin{abstract}
A novel method, based on superpotentials is proposed for obtaining
the quasi-normal modes of anti-deSitter black holes. This is
inspired by the case of the three dimensional BTZ black hole, 
where the quasi-normal modes can be obtained exactly and are
proportional to the surface gravity.
Using this
approach, the quasi-normal modes of the five dimensional Schwarzschild
anti-deSitter black hole are computed numerically. The modes 
again seem to be proportional to the surface gravity for very
small and very large black holes.
They reflect the well-known instability of small black holes in
anti-deSitter space.
\end{abstract}

\newpage
\section{Introduction}
Numerical studies of perturbations of black holes have revealed the presence of certain 
characteristic modes that govern the time evolution of the initial perturbation. These modes
seem to depend only on the black hole parameters and not on the nature of the perturbation.
This was first recognised by Vishveshvara \cite{vish} while studying perturbations of Schwarzschild
black holes. Since then, perturbations of asymptotically
flat black holes have been analysed and these characteristic modes(called quasi-normal modes) have
been found for some black holes (a detailed account can be found in \cite{nollert}, \cite{kokkotas}).

Recently, there has been an interest in quasi-normal modes of $AdS$ black holes in light of the 
$AdS/CFT$ correspondence \cite{witten}. 
There is some evidence to suggest that an off-equilibrium configuration in
the bulk $AdS$ space is related to an off equilirium state in the boundary theory.
In \cite{vakkuri}, black hole formation by collapse of a thin shell in $AdS$ space
is investigated. The shell is related by $AdS/CFT$ duality to an off-equilibrium
state in the boundary theory which evolves towards equilibrium when the shell
collapses to form a black hole.

A quasi-normal mode governs the decay in time of
a perturbation of the black hole configuration
in the bulk. It should therefore be related by $AdS/CFT$ duality to the return
of the boundary Yang-Mills theory to thermal equilibrium.
The role of ingoing modes in determining the thermalization time scale of the boundary
Yang-Mills theory was pointed out in \cite{bala}\footnote{The normalisation measure used
here is not appropriate, however, and these modes are not quasi-normal modes.}.
A numerical computation of  quasi-normal mode
frequencies for $AdS$ black holes in various dimensions has been done in \cite{horo}.

In this paper, we compute the quasi-normal modes for the $AdS$-Schwarzschild black hole in five 
dimensions using a superpotential approach.  
This is done for the perturbation of a black hole by a minimally 
coupled scalar field. Time-independent normal modes of a scalar field 
in the background of this black hole
have already been analysed \cite{gsv, witten1}.

We notice, that for the $(2+1)-d$ BTZ black hole,
exact quasi-normal mode solutions can be obtained from the Klein-Gordon equation 
as the black hole potential belongs to a class of exactly solvable
potentials, derived from a superpotential.

We propose, based on analogy with the results for the
$(2+1)-d$ BTZ black hole, and on
numerical evidence in five dimensions,
that the black hole potential can be 
described by a potential series derived from a superpotential.
The problem is then exactly solvable
and the form of the quasi-normal mode wave function is thus known. 
This is used as an ansatz to obtain
the quasi-normal modes of the $AdS_{5}$ black hole. However, the numerical values of the
mode frequencies do not agree with the earlier calculation in \cite{horo}. We comment
on this discrepancy.

The organisation of this paper is as follows: The next section is a brief review of 
quasi-normal modes and their properties, and a short summary of some numerical
approaches to computing  these modes.  

The actual computation of quasi-normal(QN) modes for the 
$AdS_{5}$-Schwarzschild black hole is described in Section 3.
First, the QN modes for the $(2+1)-d$ BTZ
black hole are evaluated. Motivated by this, a superpotential approach to finding the 
QN modes for the five-dimensional black hole is presented and the mode frequencies
obtained.

In Section 4, we discuss our results. We show that the  behaviour of the
modes computed numerically by us for small black holes is consistent
with that expected from the differential equation obeyed by the mode
solutions.
We mention salient points of two 
earlier papers \cite{bala} and \cite{horo} on the subject and comment on
the discrepancy between the numerical values obtained by us and those in \cite{horo}.
We also state  work in progress on QN modes for the
RN $AdS_{5}$ black hole, and black holes in other dimensions.
       
\section{Quasi-normal modes and their properties}
Quasi-normal modes of a black hole are characteristic modes associated
with the decay of any perturbation outside a black hole. In general,
these modes do not form a complete set, in the sense that the time
decay of a perturbation cannot be described {\em completely} in terms
of them. However, they dominate the decay at certain intermediate or
late times. To compute these modes, one studies the decay of an
initial perturbation by imposing ingoing boundary conditions at the
horizon and (for asymptotically flat black holes) outgoing boundary
conditions at infinity. This ensures that no gravitational 
wave from the horizon or infinity disturbs the initial perturbation.
Due to these boundary
conditions, the quasi-normal modes are complex. The mode wave functions
are non-normalisable in space but exponentially decay in time.
Since the exact mode wave functions are not known even for the
Schwarzschild black hole, there exist only numerical computations
of quasi-normal modes. These are also very difficult to do as it is not
easy to isolate the purely ingoing wave near the horizon that is also
outgoing at infinity. Such
a wave function  blows up at these points and could be contaminated
by a small outgoing part at the horizon or a small ingoing part at
infinity where these parts go exponentially to zero. A unique
boundary condition to isolate quasi-normal modes has been given by
Nollert \cite{nollert}. The QN modes are poles of the Green's
function corresponding to the wave equation obeyed by the
perturbing field.
Using this, the QN modes for the 
Schwarzschild and Reissner-Nordstrom black holes were evaluated
numerically \cite{chandra}, \cite{nollert1}.
WKB methods have also been used to calculate these
modes \cite{schutz} (see also \cite{nollert} for more references).

The method of continued fractions \cite{leaver} gives
very accurate results for the modes. Here, after removing the
singular parts at the horizon and infinity, the QN mode wave
function is expanded as a series. The coefficients in the series
obey a three-term recurrence relation and depend on the quasi-normal
mode frequency. The frequencies then have to satisfy
a continued fraction relation.
An interesting analytic approach to computing
the QN modes is by approximating
the black hole potential by an exactly solvable potential (the
Poschl Teller or Eckart potential) \cite{ferrari}, \cite{mashhoon}.
Ferrari and Mashhoon \cite{ferrari} have mapped the 
problem of QNMs of the Poschl-Teller potential to the bound
states of the inverse potential and computed the modes. These approximate
the fundamental QN modes of the Schwarzschild black hole well. 

For the case of asymptotically anti-deSitter black holes, the black
hole potential diverges at infinity. Therefore, as pointed out
in \cite{horo}, the quasi-normal modes are defined to be those
that are ingoing at the horizon, but fall off to zero at infinity.
It can be shown that this also agrees with the unique boundary 
condition of Nollert \cite{nollert} for obtaining quasi-normal
modes. However, due to the divergence of the $AdS$ black hole potential
at infinity, most of the methods described above to compute the QN
modes of asymptotically flat black holes cannot be used directly.
It is seen that a combination of two methods, inspired by 
the case of the $(2 + 1)-d$ black hole, yields good results.

\section{Numerical computation of quasi-normal modes}

We first consider the case of the QN modes of the non-rotating $(2 + 1)-d$ BTZ black
hole \cite{banados}. The non-rotating BTZ black hole metric  
for a spacetime with  
negative cosmological constant $\Lambda~=~-\frac{1}{l^2}$ is given by
\begin{eqnarray}
ds^2~~~~=~~~~-(N)^2~dt^2 + (N)^{-2}~dr^2 + r^2~(d\phi)^2
\label{met3}
\end{eqnarray}

with
\begin{eqnarray}
N~~=~~\sqrt{(-M~+~\frac{r^2}{l^2})}
\label{metdef3}
\end{eqnarray}
The horizon radius $r_{+} ~=~\sqrt{M} l$.
We consider a massless scalar field in the black hole background. The Klein-Gordon equation
for the scalar field is written using an ansatz (independent of the 
angular variable) for the field 
\begin{eqnarray}
\Phi = \frac{1}{\sqrt{r}}~ \chi(r)~ \exp (i \omega t)
\end{eqnarray}
and by going to the tortoise coordinate $r_{*}$ where $dr_{*} = \frac{dr}{N^2} $. The Klein-
Gordon equation is
\begin{eqnarray}
- \frac{d^2 \chi}{d r_{*}^2}~ +~ V(r) \chi ~=~ \omega^2 \chi
\label{kgeqn}
\end{eqnarray}
where
\begin{equation}
V ~=~ \frac{3 r^2}{4 l^4} ~-~ \frac{M}{2 l^2} ~-~ \frac{M^2}{4 r^2}
\end{equation}
Since $r = - r_{+} \coth (\alpha r_{*})$ where 
\begin{eqnarray}
\alpha ~=~ \frac{1}{2}~\frac{d(N^2)}{dr}(r_{+}) ~=~ \frac{\sqrt{M}}{l}
\label{sgrav}
\end{eqnarray}
\begin{eqnarray}
V ~=~ \frac{M}{4 l^2}~  (\frac{3}{(\sinh (\alpha r_{*}))^2} ~~+~~ 
\frac{1}{(\cosh (\alpha r_{*}))^2}) 
\label{pot3}
\end{eqnarray}
$\alpha$ is the surface gravity of the black hole and is equal to $2\pi T$
where $T$ is the Hawking temperature.

This potential can be obtained from a superpotential \cite{khare}
of the form
\begin{eqnarray}
W ~=~ A~\coth (\alpha r_{*}) ~~+~~ B~ \tanh (\alpha r_{*})
\label{superpot3}
\end{eqnarray}
by
\begin{eqnarray}
V ~=~ W^2 ~-~ (W)' ~-~ A^2 ~-~ B^2 ~-~ 2 A B
\label{dpot3}
\end{eqnarray}
Here, from (\ref{pot3}), 
\begin{eqnarray}
A ~=~ - \alpha (1/2 \pm 1) \\
B ~=~ - \frac{\alpha}{2} 
\label{abdef}
\end{eqnarray}

 As discussed in \cite{khare}, 
the lowest energy state  for the potential $W^2 - (W)'$ has energy zero, so
the lowest energy for the potential $V$ is $- (A + B)^2 $. The
wave function corresponding to this state is

\begin{eqnarray}
\chi ~=~ \exp (- \int (W)) ~~~=~~~ (\sinh (\alpha r_{*}))^{- A/(\alpha)}
~~(\cosh (\alpha r_{*}))^{- B/(\alpha)}
\label{wf3}
\end{eqnarray}
For any of the two values of $A$ and $B$ in (\ref{abdef}), 
the wave function (\ref{wf3}) is not normalisable.
It blows up either at the horizon or at $r = \infty$ (i.e $r_{*} = 0$). We are interested in the
quasi-normal mode solution that blows up at the horizon and falls off to zero at infinity. This
corresponds to $A = - (3/2) \alpha$.  Also, $B = -\frac{\alpha}{2} $. The lowest
energy (formally, since the wave function corresponding to this is non-normalisable) is then 
\begin{equation}
E ~=~ -4 (\alpha)^2
\end{equation}
and the lowest quasi-normal mode $\omega = \sqrt{E}$. Therefore,

\begin{equation}
\omega ~~=~~ - 2~i~\alpha
\end{equation}

Thus, for a given horizon radius $r_{+} = \sqrt{M} l$, we can
calculate $\alpha$ taking $l=1$. Then we can read off the value
of the lowest quasi-normal mode. 

The case of the quasi-normal modes of the $AdS_{5}$-Schwarzschild
black hole is not so simple as the black hole potential is more
complicated than the BTZ case.

The metric for this black hole is 
\begin{eqnarray}
ds^2~~~~=~~~~-(N)^2~dt^2 + (N)^{-2}~dr^2 + r^2~(d\Omega_{3})^2
\label{met5}
\end{eqnarray}

with
\begin{eqnarray}
N~~=~~\sqrt{1 ~+~ (r^2)/(l^2) ~-~ (r_{0}/r)^2}
\label{metdef5}
\end{eqnarray}
where 
the black hole mass $M$ is
\begin{equation}
M = \frac{3 A_{3} r_{0}^2}{16 \pi G_{5}}
\end{equation}
and $A_{3}$ is the area of a unit 3-sphere.
The Klein-Gordon equation for the scalar field in the background of
this black hole can be written, as before, as a potential problem. Using
the ansatz for the field
\begin{eqnarray}
\Phi ~=~ \left( \frac{1}{r} \right)^{3/2}~\chi (r) ~ \exp (i \omega t)
\label{phi5}
\end{eqnarray}
and by changing to the tortoise coordinate
given by $dr_{*}= \frac{dr}{N^2}$, the Klein Gordon equation is

\begin{eqnarray}
-~ \frac{d^2 \chi}{dr_{*}^2} ~~+~~V(r)~\chi~~=~~\omega^2 ~\chi 
\label{eqn5}
\end{eqnarray}

where 
\begin{eqnarray}
V(r) ~~=~~\left( \frac{15}{4} + \frac{3}{4 r^2} + \frac{9 r_{0}^2}{4 r^4} \right)
\left( 1 +  r^2 - (\frac{r_{0}}{r})^2 \right)
\label{pot5}
\end{eqnarray}

Here, for simplicity, we have taken $l=1$. We have also not considered an 
angular dependence for the field, which is a simple generalisation of the 
following analysis.

If the potential could be written as an exactly solvable potential
in the $r_{*}$ coordinate, then the quasi-normal modes could be
obtained easily from the `lowest' energy, as in the BTZ case.
As mentioned in the previous section, the 
fundamental QN modes of the Schwarzschild
black hole were closely approximated by the QN modes of the Poschl-Teller
potential. This is also an exactly solvable potential derivable from
a superpotential \cite{khare}.

We propose that the potential can be written as a series, derived from a 
superpotential of the form
\begin{eqnarray}
W ~=~ \sum_{n=1}^{\infty} A_{n} \coth (n \alpha r_{*}) 
\label{superpot5}
\end{eqnarray}
where as in (\ref{sgrav}), $\alpha$ is the black hole surface gravity,
\begin{eqnarray}
\alpha ~=~(1/2)~\frac{d(N^2)}{dr}(r_{+}) ~=~r_{+}~+~\frac{r_{0}^2}{r_{+}^3}
\end{eqnarray}
As before, $\alpha = 2\pi T$, where $T$ is the Hawking temperature.

The potential derived from this series as in (\ref{dpot3}) is

\begin{eqnarray}
V_{s} ~=~ \sum_{n=1}^{\infty}\frac{(A_{n}^2 + n A_{n} \alpha)}{\sinh^2 (n \alpha r_{*})} ~+~
\sum_{n,m=1, ~n \neq m}^{\infty} \frac{A_{n} A_{m}}{\tanh (n \alpha r_{*}) \tanh (m \alpha r_{*})}
\label{potser5}
\end{eqnarray}
It has lowest  energy
\begin{equation}
E ~=~ - (\sum_{n=1}^{\infty} A_{n})^2
\end{equation}
The form of the potential is such that it goes exponentially to zero in $r_{*}$ as
$r \rightarrow r_{+}$ and blows up as $1/r_{*}^2$ as $r_{*} \rightarrow 0$ (or
$r \rightarrow \infty$). This reproduces the behaviour of the black hole potential. We
propose that this potential series is exactly equal to the black hole potential (\ref{pot5}).
Terminating this potential series at finite order would then mean an approximation to the 
black hole potential in the spirit of the Poschl-Teller method for asymptotically flat 
black holes.
The wave function corresponding to the lowest energy state is
\begin{eqnarray}
\chi ~=~ \exp (- \int (W)) ~~~=~~~ \prod_{n=1}^{\infty} (\sinh (n \alpha r_{*}))^{- A_{n}/(n \alpha)}
\label{wf5}
\end{eqnarray}
It is non-normalisable, as in the $(2 + 1)-d$ case.
In order to calculate the quasi-normal modes for the $AdS$-Schwarzschild black hole,
we use (\ref{wf5}) as an ansatz in the equation (\ref{eqn5}). However, we truncate the 
product in the ansatz upto some finite order. This is equivalent to truncating the 
potential series.
We first write
\begin{eqnarray}
\chi ~~=~~ \psi~ \exp (-i \omega r_{*})
\label{chi}
\end{eqnarray}
This isolates the non-normalisable part of the wave function.
We can rewrite (\ref{eqn5}) as an equation for $\psi$ as
\begin{eqnarray}
-~\frac{d^2 \psi}{dr_{*}^2} ~~+~~2 i \omega \frac{d\psi}{dr_{*}} ~~+~~ V(r) \psi ~~=~~0
\label{eq5}
\end{eqnarray}
Here, $V$ is the black hole potential (\ref{pot5}). We now try to solve for $\omega$ by taking the
ansatz

\begin{eqnarray}
\psi ~~=~~ \prod_{n=1}^{N}(1~-~\exp (2 n \alpha r_{*}))^{- \frac{A_{n}}{n\alpha}}
\label{psi5}
\end{eqnarray}
This ansatz is obtained from (\ref{wf5}) which was an ansatz for $\chi$. Thus, $N$ represents
the order upto which $V_{s}$ has been truncated.

The ansatz for $\psi$ is substituted into (\ref{eq5}) and the l.h.s of (\ref{eq5}) is expanded
as a series in $(x - x_{+})$ where 
\begin{eqnarray}
x~=~\frac{1}{r} \nonumber \\ 
x_{+}~=~\frac{1}{r_{+}}
\label{xdef}
\end{eqnarray}
From (\ref{eq5}), each term of the series is to be equated to zero. 
We first divide (\ref{eq5}) by $\psi$. Then, we have
\begin{eqnarray}
-~4 (\sum_{m=1}^{N} A_{m} K_{m})^2 ~-~ 4~\alpha~(\sum_{m=1}^{N} m A_{m} K_{m}) ~-~ \nonumber \\
4~\alpha~(\sum_{m=1}^{N} m A_{m} K_{m}^2 ) ~+~4~i\omega~(\sum_{m=1}^{N} A_{m} K_{m}) ~+~V~~=~~0
\label{receqn}
\end{eqnarray}

where
\begin{equation}
K_{m}~~=~~ \frac{\exp (2m\alpha r_{*})}{1~-~\exp ((2m\alpha r_{*})}
\end{equation}
Expanding each of the terms in the l.h.s of (\ref{receqn}),
\begin{eqnarray}
K_{m} ~~=~~ \sum_{n=0} c_{m}(n)~(x - x_{+})^n \\
K_{m}^2 ~~=~~ \sum_{n=0} d_{m}(n)~(x - x_{+})^n \\ 
V ~~=~~\sum_{n=0} V_{n}~(x - x_{+})^n \nonumber \\ 
\end{eqnarray}
Substituting these expansions into (\ref{receqn}), and equating the term of
order $(x - x_{+})$ in the l.h.s to zero, we obtain

\begin{equation}
A_{1}~~=~~-~\frac{\alpha (\frac{15}{4}~+~\frac{3 x_{+}^2}{4}~+~\frac{9r_{0}^2 x_{+}^4}{4})}
{2 (i\omega~-~\alpha) c_{1}(1)}
\end{equation}
Similarly, equating the term of order $(x - x_{+})^N$ in the l.h.s to zero, we obtain 
a recursion relation for $A_{N}$. 

\begin{eqnarray}
A_{N}~~=~~ \frac{- \alpha (\sum_{m=1}^{N-1} m A_{m}c_{m}(N))}{(N\alpha~-~i\omega)c_{N}(N)} 
~-~ \frac{\alpha (\sum_{m=1}^{N-1} m A_{m} d_{m}(N))}{(N\alpha~-~i\omega)c_{N}(N)} \nonumber \\
~-~\frac{(\sum_{m,n=1}^{N-1}~\sum_{N1,N2;N1+N2=N} A_{m} A_{n} c_{m}(N1) c_{n}(N2))}{(N\alpha~-~i\omega)c_{N}(N)}  \nonumber \\
~+~i\omega~\frac{(\sum_{m=1}^{N-1} A_{m}c_{m}(N))}{(N\alpha~-~i\omega)c_{N}(N)}  
~+~\frac{V_{N}}{4(N\alpha~-~i\omega)c_{N}(N)} \nonumber \\
\label{recur}
\end{eqnarray}

The $A_{n}$ coefficients are functions of $\omega$ and of the black hole parameters.
The recursion relation for the $A_{n}$ is complicated, unlike the case of the
method of partial fractions \cite{leaver} applied to asymptotically flat black 
holes. There, an ansatz was taken for the wave function (removing its singular
part) and it was expanded as a series whose coefficients obeyed a simple
three-term recursion relation.
From (\ref{eq5}),
as $r_{*} \rightarrow 0$ (i.e as $r \rightarrow \infty$), the two kinds of solutions are
$\psi \sim r_{*}^\frac{-3}{2}$ and $\psi \sim r_{*}^\frac{5}{2}$. The first solution is not
normalisable. We want the quasi-normal mode solution to vanish as $r_{*} \rightarrow 0$. So we
choose the second solution. But from our ansatz (\ref{psi5}) for the solution, this implies a relation
between the coefficients $A_{n}$. More precisely,
\begin{eqnarray}
\sum_{n=1}^{N} \frac{A_{n}}{n\alpha} ~~=~~ -~5/2
\label{asy5}
\end{eqnarray}

Since the $A_{n}$ coefficients are functions of $\omega$, this relation gives the value of the 
lowest quasi-normal mode as a function of the black hole parameters. This would then be an
approximation to the actual QN mode of the black hole at order $N$. 
Thus,
the asymptotic behaviour of the actual solution is {\em explicitly}
demanded of the ansatz. This is {\em used} to obtain the mode frequencies
from the relation (\ref{asy5}).

We have used this method to calculate the QN modes at various orders $N$. We find
that as we increase $N$, there is a convergence in the mode frequency.
The mode frequencies are given in Table 1. They have been calculated at order
$N=18$. Beyond this order, as (\ref{asy5}) is complicated, it becomes 
computationally time-consuming to 
find its roots.
However, as there is a clear convergence in the mode frequency as 
one increases the order
from $N=1$ and as
the  percentage difference between the real and imaginary parts of 
the mode 
frequency at order $N=18$ and $N=17$
is about 1\% (for $Re\omega$, it is 1.26\%, and for $Im\omega$, it is 1.09\%),
these numbers should be a good approximation to the QN modes of the
black hole (more discussion on results is given in the next section).
\begin{figure}
\begin{center}
\hspace{0.5cm}
\input{table.eepic}
\vskip 1cm
\begin{tabular}{ll}
{\bf Table 1.}
\end{tabular}
\end{center}
\end{figure}

\section{Discussion of results}
The lowest quasi-normal mode frequencies for different horizon radii $r_{+}$ are
given in Table 1. As mentioned in the last section, there is a convergence in the
mode frequency as the order $N$ is increased. The convergence curve for
$r_{+}=10$ is given in Fig.1. The rate of convergence does not seem to depend
on the value of the horizon radius $r_{+}$.
\begin{figure}[t]
\centerline{\epsfxsize 5in
\epsfbox{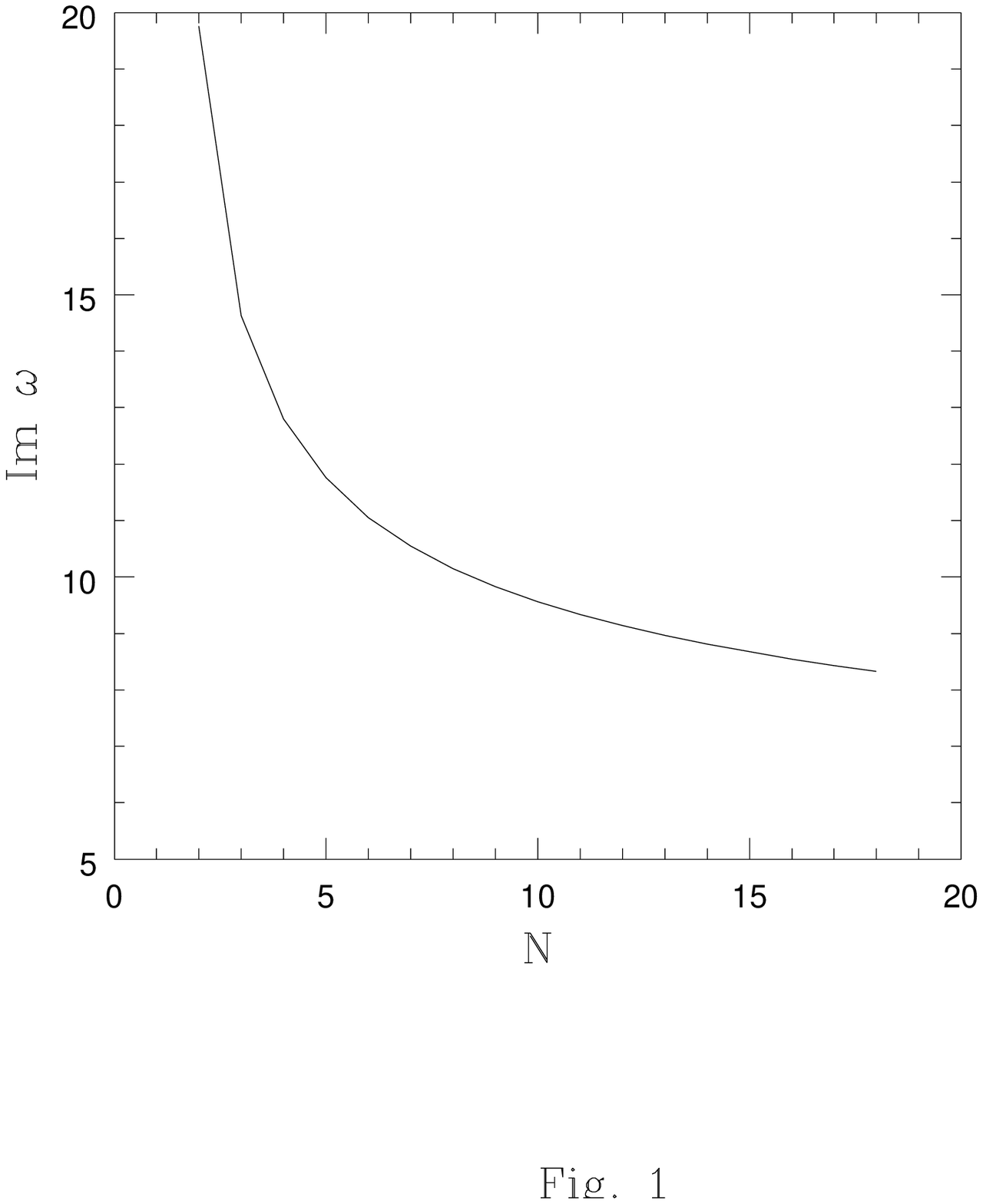}}
\begin{tabular}{ll}
{\bf Figure 1.}  
{\sl Convergence in $Im(\omega)$ as order N is increased, for $r_{+}=10$}
\end{tabular}
\end{figure}

The real and imaginary parts of the mode frequencies are plotted as a function 
of $r_{+}$ for large $r_{+}$ in Fig.2a and Fig.2b. 
\begin{figure}[t]
\centerline{\epsfxsize 5in
\epsfbox{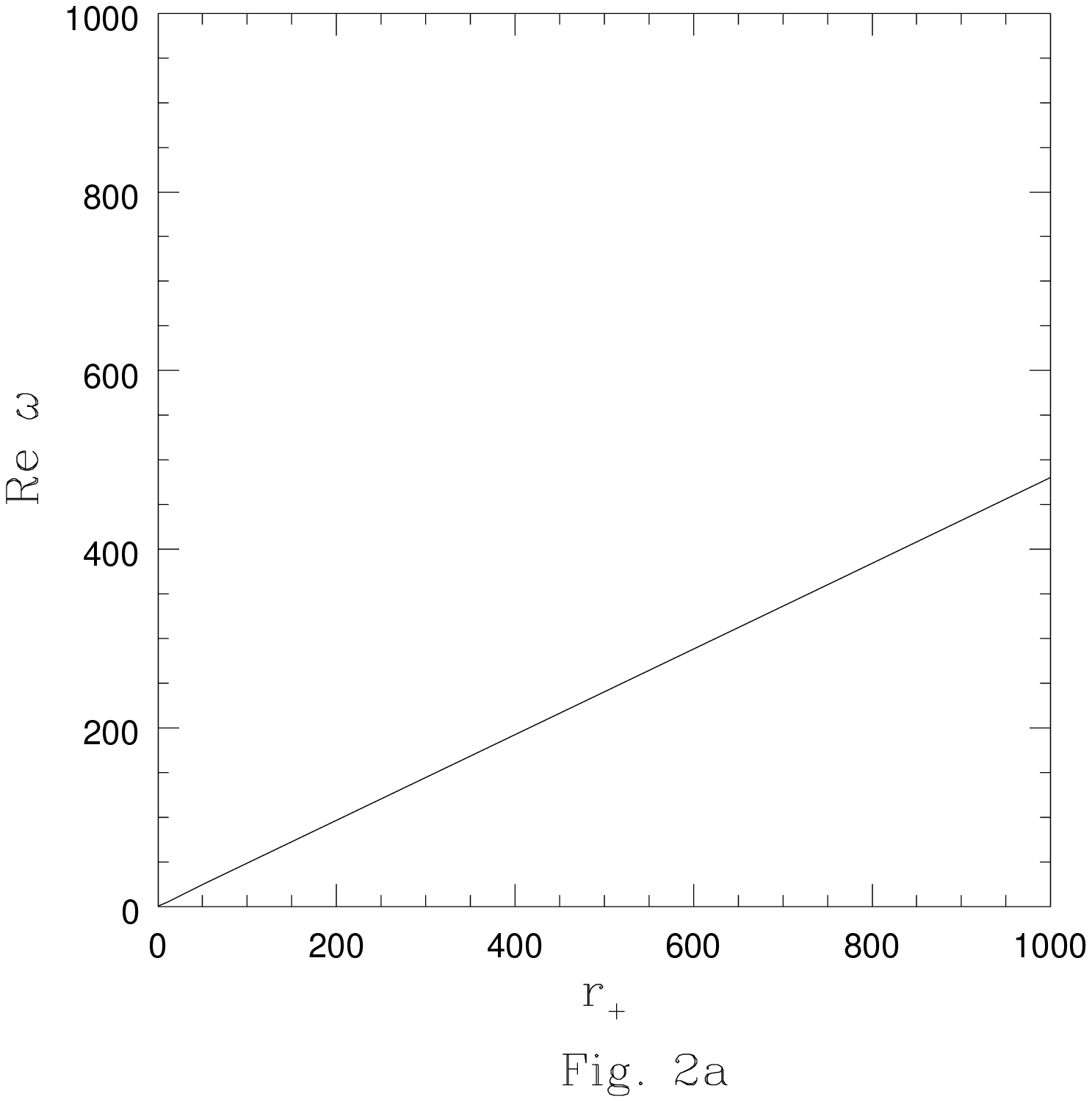}}
\begin{tabular}{ll}
{\bf Figure 2a.} 
{\sl $Re(\omega)$ as a function of $r_{+}$ for large $r_{+}$}
\end{tabular}
\end{figure}

\begin{figure}[t]
\centerline{\epsfxsize 5in
\epsfbox{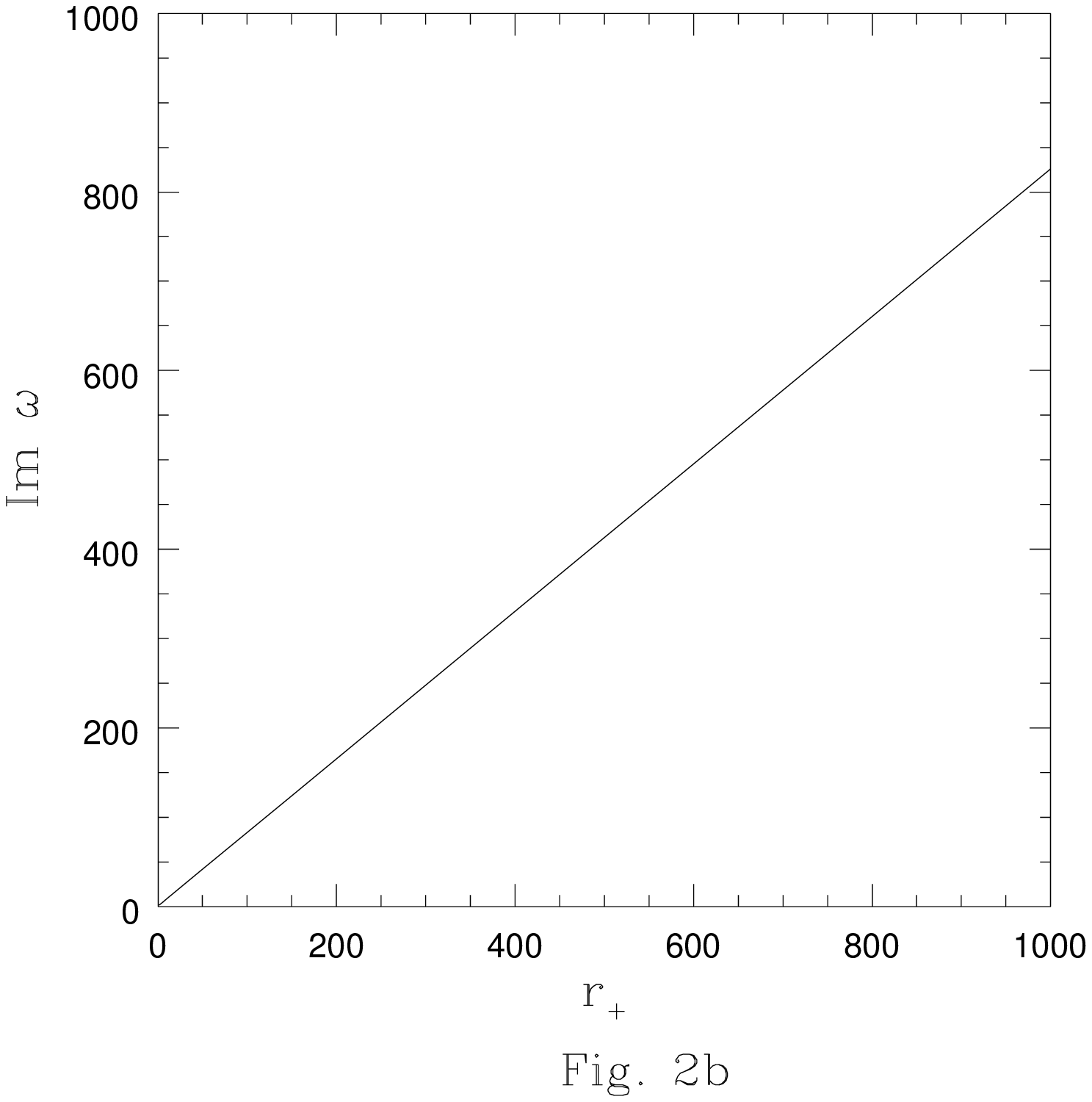}}
\begin{tabular}{ll}
{\bf Figure 2b.} 
{\sl $Im(\omega)$ as a function of $r_{+}$ for large $r_{+}$}
\end{tabular}
\end{figure}

It is seen that the both the imaginary and real parts of the mode frequency
are proportional to $r_{+}$ for {\em large} $r_{+}$. The real and imaginary parts
of the mode frequency can also be plotted as a function of the surface gravity
$\alpha$. 
It is seen that for {\em both} small and large $r_{+}$, the real and
imaginary parts
of the frequency are approximately proportional to $\alpha$.
$Re(\omega) \sim 0.24~\alpha$ and $Im(\omega) \sim 0.41~\alpha$ provides a 
good fit for the data. 
In the case of the QN modes where the modes could be obtained exactly, we saw
that they were proportional to the surface gravity. Our numerical results seem
to suggest that this may be true even for the five dimensional black hole -
at least for very small and very large black holes.
The surface gravity $\alpha ~=~ (2r_{+}~+~\frac{1}{r_{+}})$. Therefore, $\alpha$
is large for very small and very large black holes. We have verified numerically that
for very small black holes, the mode frequencies are very large, and $\omega \sim \frac{1}{r_{+}}$.
This behaviour of the mode frequency is expected - as can be seen from the differential 
equation (\ref{eq5}) on changing to the 
inverse radial coordinate $x$ given by (\ref{xdef}). Then, the 
differential equation in these coordinates is
\begin{eqnarray}
(- r_{0}^{2} x^6 + x^4 + x^2) \frac{d^2 \psi}{dx^2}~+~
(2 x^3 - 4 r_{0}^{2} x^5 + 2 i \omega x^2)\frac{d \psi}{dx}~-~(\frac{15}{4} + \frac{3 x^2}{4} +
\frac{9 r_{0}^{2} x^4}{4}) \psi ~=~ 0
\label{xeqn}
\end{eqnarray}
Now, we scale $x$ as $x~=~q~x_{+}$ (where $x_{+}= \frac{1}{r_{+}}$).
Now, (\ref{xeqn}) is
\begin{eqnarray}
( - x_{+}^{2} q^6 - q^6 + x_{+}^{2} q^4 + q^2) \frac{d^2 \psi}{dq^2}~+~ \nonumber \\
(2 x_{+}^2 q^3 - 4 x_{+}^2 q^5 - 4 q^5 + 2 i \omega x_{+} q^2) \frac{d \psi}{dq} ~-~\nonumber \\
(\frac{15}{4} + \frac{3 x_{+}^{2} q^2}{4} + \frac{9 x_{+}^{2} q^4}{4} + \frac{9 q^4}{4})\psi~=~0
\label{xeqn1}
\end{eqnarray}
Then we see that for very large black holes (small $x_{+}$ approximation),
$x_{+}$ can be scaled away from (\ref{xeqn1}) near the horizon in this approximation
provided $\omega~=~\frac{C}{x_{+}}$ where C is a constant
independent of $x_{+}$. It can also be checked that near the horizon, there are no
solutions to the scaled equation with $C~=~0$, except the trivial solution. This shows that
$\omega$ is indeed proportional to $r_{+}$ (i.e $\frac{1}{x_{+}}$)
for large black holes.

For very small black holes (i.e in the large $x_{+}$ approximation), again $x_{+}$ can
be scaled away from (\ref{xeqn1}) near the horizon provided $\omega~=~ D~x_{+}$ where D is a constant
independent of $x_{+}$. Here too, it can be checked that near the horizon, there are
no solutions to (\ref{xeqn1}) in the large $x_{+}$ approximation with $D~=~0$ (except the 
trivial solution). This implies that for small black holes, $\omega$ is proportional to
$\frac{1}{r_{+}}$ (i.e to $x_{+}$). This is also just a reflection of the fact that a very small 
$AdS$-Schwarzschild black hole has negative specific heat and actually resembles a 
Schwarzschild black hole. 

\begin{figure}[t]   
\centerline{\epsfxsize 5in 
\epsfbox{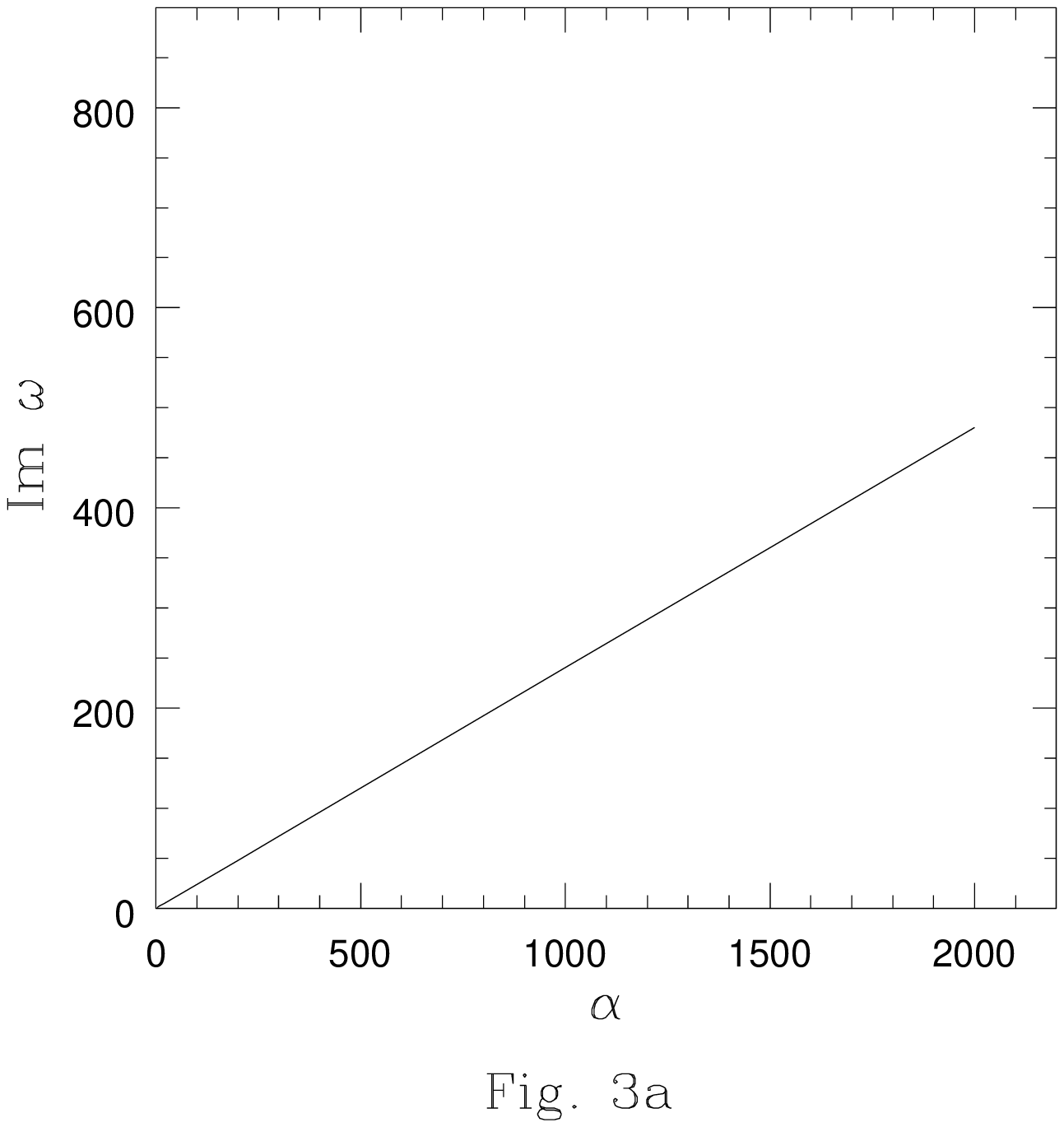}} 
\begin{tabular}{ll} 
{\bf Figure 3a.}  
{\sl Re($\omega$) as a function of $\alpha$} 
\end{tabular}   
\end{figure}

\begin{figure}[t]      
\centerline{\epsfxsize 5in  
\epsfbox{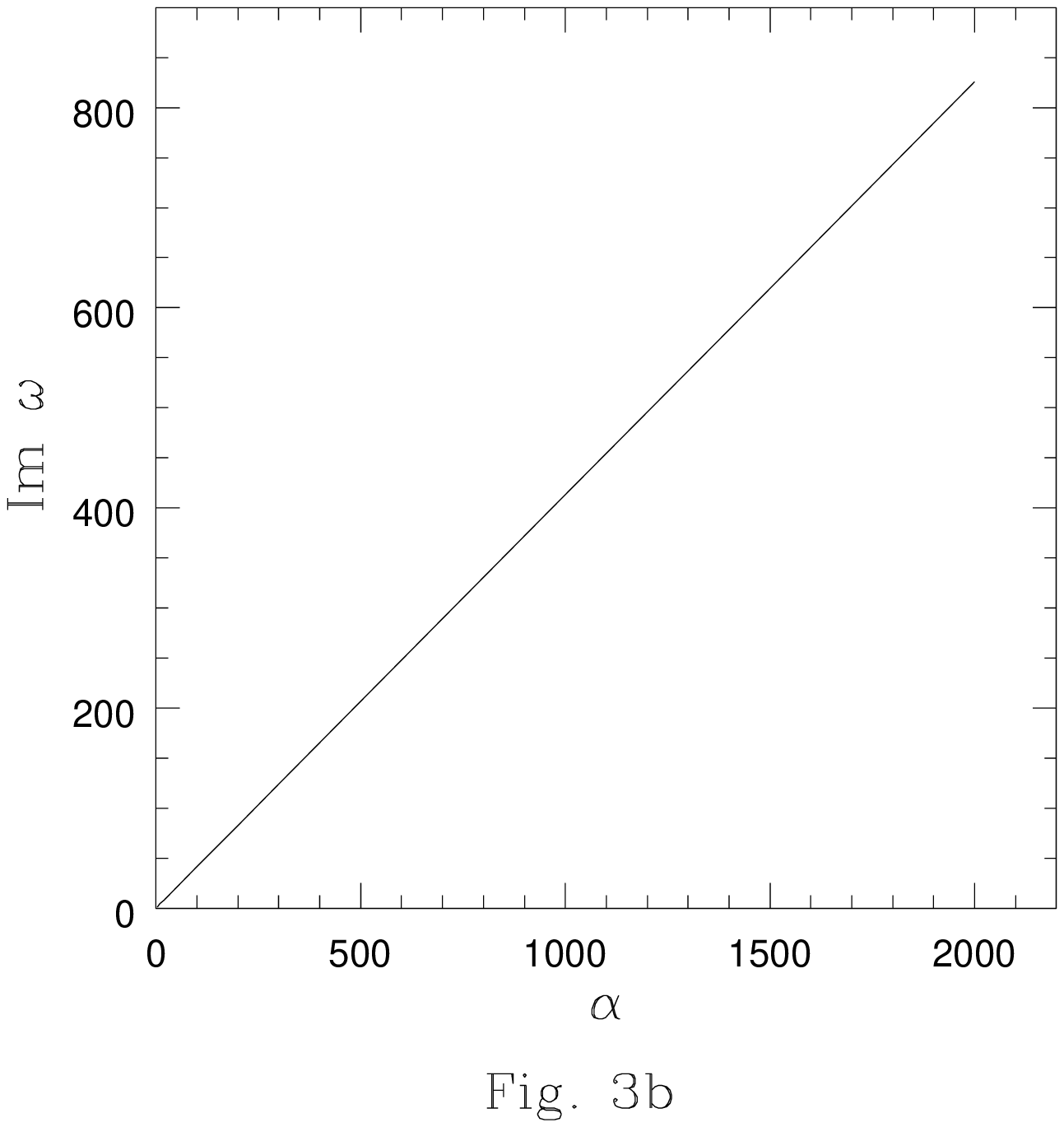}}  
\begin{tabular}{ll}  
{\bf Figure 3b.}   
{\sl Im($\omega$) as a function of $\alpha$}  
\end{tabular}      
\end{figure}

We now examine the previous numerical work on   
QN modes of $AdS$ black holes in four, five and seven dimensions. 

The numerical results obtained in \cite{horo} for the five dimensional
black hole do not agree with our values for the QN modes.
There, the solution to (\ref{eq5}) is expanded as a series around the horizon in the 
inverse radial coordinate $x=1/r$. Using this, the l.h.s of (\ref{eq5}) is 
expanded as a series around the horizon, and a recursion relation is obtained for 
the coefficients in the expansion of the solution. The coefficients $a_{n}$ are
functions of the black hole parameters and the mode frequency $\omega$. Therefore,
\begin{eqnarray}
\psi ~~=~~\sum_{n} a_{n}(\omega)~(x - x_{+})^{n}
\label{horeq1}
\end{eqnarray}
The mode frequencies are then the 
roots of the equation obtained by setting the series to zero at $x=0$. In actual
computation, the series is truncated, and the roots obtained. However, the mode frequencies
seem to go to zero as $r_{+} \rightarrow 0 $, which is not the behaviour expected from the
differential equation. 

\pagebreak

A numerical computation of QN modes is tricky because of the 
nature of the boundary conditions on the mode solution at both the horizon and at infinity.
The mode, which is ingoing (and not normalisable) at the horizon, could be contaminated by an
outgoing component which goes to zero there.
At the other boundary, the solution is required to go to zero.
In a numerical truncation, 
it could also be contaminated as $x \rightarrow 0$ (i.e as $r \rightarrow \infty$) 
by the solution that blows up at this end. Therefore, in a numerical computation, it is essential
to ensure the correct ingoing behaviour near the horizon and the correct {\em asymptotic}
behaviour of the solution that goes to zero as $r \rightarrow \infty$.
In our method, similar to the continued fraction method for asymptotically
flat black holes, we  
have a {\em specific} ansatz for the wave function where the 
behaviour at {\em both} boundaries is explicitly present in the form of the ansatz.
The form of our ansatz
ensures that there is no contamination from the outgoing part at the horizon. At the other end
as $x \rightarrow 0$ (i.e as $r \rightarrow \infty$), 
we {\em demand} that the solution must fall off as $x^{\frac{5}{2}}$ at every order.
This {\em ensures} that there is no contamination from the solution that blows up at this end.

We have presented a new approach to computing the quasi-normal modes of 
$AdS$ black holes. 
The novel feature of this method was an ansatz for the 
QN mode wave function which was derived from a superpotential.
This was made in analogy with the case of the
three dimensional BTZ black hole where the modes can be obtained 
exactly and the wave function is derived from a superpotential.
The BTZ QN modes were proportional to the surface gravity.
The modes obtained by us numerically for the five dimensional
black hole are also approximately proportional to the surface
gravity. 
More importantly, the modes obtained by us numerically are proportional
to the inverse of the horizon radius for small black holes, reflecting the well-known
fact that these black holes are unstable. We have also shown, from some scaling
properties of the differential equation obeyed by the mode solutions, that this
is indeed to be expected for small black holes.
Work is in progress to compute
the QN modes for AdS black holes in four and seven dimensions, and
also the corresponding Reissner-Nordstrom black holes using
this approach. Some preliminary work on the five dimensional
RN $AdS$ black hole seems to suggest that the QN mode
increases with the charge of the black hole. 

\vskip 1cm
We would like to thank G.T.Horowitz and V.Hubeny for useful comments about the results in our
paper.

\pagebreak

\end{document}

%% file: table.eepic
\setlength{\unitlength}{0.0125in}
\begingroup\makeatletter\ifx\SetFigFont\undefined
\def\x#1#2#3#4#5#6#7\relax{\def\x{#1#2#3#4#5#6}}%
\expandafter\x\fmtname xxxxxx\relax \def\y{splain}%
\ifx\x\y   
\gdef\SetFigFont#1#2#3{%
  \ifnum #1<17\tiny\else \ifnum #1<20\small\else
  \ifnum #1<24\normalsize\else \ifnum #1<29\large\else
  \ifnum #1<34\Large\else \ifnum #1<41\LARGE\else
     \huge\fi\fi\fi\fi\fi\fi
  \csname #3\endcsname}%
\else
\gdef\SetFigFont#1#2#3{\begingroup
  \count@#1\relax \ifnum 25<\count@\count@25\fi
  \def\x{\endgroup\@setsize\SetFigFont{#2pt}}%
  \expandafter\x
    \csname \romannumeral\the\count@ pt\expandafter\endcsname
    \csname @\romannumeral\the\count@ pt\endcsname
  \csname #3\endcsname}%
\fi
\fi\endgroup
\begin{picture}(241,276)(0,-10)
\path(1,261)(241,261)
\path(1,221)(241,221)
\path(1,121)(241,121)
\path(1,101)(241,101)
\path(1,61)(241,61)
\path(1,41)(241,41)
\path(1,21)(241,21)
\path(1,81)(241,81)
\path(1,161)(241,161)
\path(0,141)(240,141)
\path(0,260)(0,0)
\path(73,260)(73,1)
\path(158,261)(158,1)
\path(241,260)(241,1)
\path(1,1)(241,1)
\path(1,201)(241,201)
\path(1,181)(241,181)
\put(96,241){\makebox(0,0)[lb]{\smash{{{\SetFigFont{12}{14.4}{rm}Re($\omega$)}}}}}
\put(167,241){\makebox(0,0)[lb]{\smash{{{\SetFigFont{12}{14.4}{rm}Im($\omega$)}}}}}
\put(101,206){\makebox(0,0)[lb]{\smash{{{\SetFigFont{12}{14.4}{rm}0.6948}}}}}
\put(101,166){\makebox(0,0)[lb]{\smash{{{\SetFigFont{12}{14.4}{rm}2.4462}}}}}
\put(101,146){\makebox(0,0)[lb]{\smash{{{\SetFigFont{12}{14.4}{rm}4.8249}}}}}
\put(96,126){\makebox(0,0)[lb]{\smash{{{\SetFigFont{12}{14.4}{rm}24.0159}}}}}
\put(96,106){\makebox(0,0)[lb]{\smash{{{\SetFigFont{12}{14.4}{rm}48.0251}}}}}
\put(96,86){\makebox(0,0)[lb]{\smash{{{\SetFigFont{12}{14.4}{rm}72.0358}}}}}
\put(91,66){\makebox(0,0)[lb]{\smash{{{\SetFigFont{12}{14.4}{rm}240.1150}}}}}
\put(91,46){\makebox(0,0)[lb]{\smash{{{\SetFigFont{12}{14.4}{rm}360.1720}}}}}
\put(91,26){\makebox(0,0)[lb]{\smash{{{\SetFigFont{12}{14.4}{rm}480.2290}}}}}
\put(101,186){\makebox(0,0)[lb]{\smash{{{\SetFigFont{12}{14.4}{rm}1.0713}}}}}
\put(27,146){\makebox(0,0)[lb]{\smash{{{\SetFigFont{14}{16.8}{rm}10}}}}}
\put(14,105){\makebox(0,0)[lb]{\smash{{{\SetFigFont{14}{16.8}{rm}   100}}}}}
\put(170,26){\makebox(0,0)[lb]{\smash{{{\SetFigFont{12}{14.4}{rm}826.0980}}}}}
\put(169,45){\makebox(0,0)[lb]{\smash{{{\SetFigFont{12}{14.4}{rm}619.5740}}}}}
\put(169,65){\makebox(0,0)[lb]{\smash{{{\SetFigFont{12}{14.4}{rm}413.0500}}}}}
\put(168,86){\makebox(0,0)[lb]{\smash{{{\SetFigFont{12}{14.4}{rm}123.9190}}}}}
\put(175,105){\makebox(0,0)[lb]{\smash{{{\SetFigFont{12}{14.4}{rm}82.6165}}}}}
\put(175,125){\makebox(0,0)[lb]{\smash{{{\SetFigFont{12}{14.4}{rm}41.3183}}}}}
\put(181,146){\makebox(0,0)[lb]{\smash{{{\SetFigFont{12}{14.4}{rm}8.3279}}}}}
\put(180,165){\makebox(0,0)[lb]{\smash{{{\SetFigFont{12}{14.4}{rm}4.2642}}}}}
\put(180,206){\makebox(0,0)[lb]{\smash{{{\SetFigFont{12}{14.4}{rm}1.4648}}}}}
\put(178,186){\makebox(0,0)[lb]{\smash{{{\SetFigFont{12}{14.4}{rm}1.9817}}}}}
\put(11,241){\makebox(0,0)[lb]{\smash{{{\SetFigFont{12}{14.4}{rm}Radius $r_{+}$}}}}}
\put(28,165){\makebox(0,0)[lb]{\smash{{{\SetFigFont{14}{16.8}{rm}  5}}}}}
\put(26,124){\makebox(0,0)[lb]{\smash{{{\SetFigFont{14}{16.8}{rm}50}}}}}
\put(15,66){\makebox(0,0)[lb]{\smash{{{\SetFigFont{14}{16.8}{rm} 500}}}}}
\put(8,25){\makebox(0,0)[lb]{\smash{{{\SetFigFont{14}{16.8}{rm} 1000}}}}}
\put(16,46){\makebox(0,0)[lb]{\smash{{{\SetFigFont{14}{16.8}{rm} 750}}}}}
\put(20,85){\makebox(0,0)[lb]{\smash{{{\SetFigFont{14}{16.8}{rm}150}}}}}
\put(35,206){\makebox(0,0)[lb]{\smash{{{\SetFigFont{14}{16.8}{rm}1}}}}}
\put(35,185){\makebox(0,0)[lb]{\smash{{{\SetFigFont{14}{16.8}{rm}2}}}}}
\end{picture}